\documentclass[twocolumn,showpacs,amsmath,amssymb,aps,pra,floatfix,nofootinbib,superscriptaddress]{revtex4}

\usepackage{graphicx,graphics,times,color}
\usepackage{bm}
\usepackage{epsfig}
\usepackage{epstopdf}
\usepackage{longtable}
\usepackage{color}

\def\be{\begin{equation}}
\def\ee{\end{equation}}
\def\ba{\begin{eqnarray}}
\def\ea{\end{eqnarray}}
\def\la{\langle}
\def\ra{\rangle}

\begin{document}

\title{Adiabatic many-body state preparation and information transfer in quantum dot arrays}

\author{Umer Farooq}
\affiliation{School of Science and Technology, Physics Division, University of Camerino,
62032 Camerino, Italy}
\affiliation{INFN-Sezione di Perugia, Via A. Pascoli, I-06123 Perugia, Italy}

\author{Abolfazl Bayat}
\affiliation{Department of Physics and Astronomy, University College London, Gower St., London WC1E 6BT, United Kingdom}

\author{Stefano Mancini}
\affiliation{School of Science and Technology, Physics Division, University of Camerino,
62032 Camerino, Italy}
\affiliation{INFN-Sezione di Perugia, Via A. Pascoli, I-06123 Perugia, Italy}

\author{Sougato Bose}
\affiliation{Department of Physics and Astronomy, University College London, Gower St., London WC1E 6BT, United Kingdom}

\date{\today}

\begin{abstract}
Quantum simulation of many-body systems are one of the most interesting tasks of quantum technology. Among them is the preparation of a many-body system in its ground state when the vanishing energy gap makes the cooling mechanisms ineffective. Adiabatic theorem, as an alternative to cooling, can be exploited for driving the many-body system to its ground state. In this paper, we study two most common	disorders in quantum dot arrays, namely exchange coupling fluctuations and hyperfine interaction, in adiabatically preparation of ground state in such systems. We show that the adiabatic ground state preparation is highly robust against those disorder effects making it good analog simulator. Moreover, we also study the adiabatic quantum information transfer, using singlet-triplet states, across a spin chain. In contrast to ground state preparation the transfer mechanism is highly affected by disorder and in particular,  the hyperfine interaction is very destructive for the performance. This suggests that for communication tasks across such arrays adiabatic evolution is not as effective and quantum quenches could be preferable.
\end{abstract}


\maketitle

\section{Introduction}

Recent progress in experimental realization of quantum many-body systems have made quantum simulators very desirable. In particular, simulating the ground state of certain many-body systems are highly important for both condensed matter physics and quantum technology. Cold atoms in optical lattices have been used for observing the quantum phase transition of superfluid to Mott insulator \cite{Mott-insulator-boson}. Recent achievement of single site addressing \cite{Bloch-single-site} made it possible to simulate spin wave \cite{Bloch-spin-wave}  and magnon propagation \cite{Bloch-magnon} in one dimensional arrays of cold atoms in optical lattices. While experimental achievements in optical lattices are very promising an analogues quantum simulator in solid state physics is at the edge of realization \cite{Smith-QDarray}. There are two different physical setups in solid state physics which may be used as quantum simulators in a near future: (i) an array of gated quantum dots  \cite{Smith-QDarray}; (ii) a chain of dopants, such as Phosphorus,  in silicon \cite{Simmons-RMP}. Although these systems may realize the same physical Hamiltonians as cold atoms they have important differences as well. For instance, these systems are charged particle fermions which do not exist in cold atoms and unlike cold atom systems, which are almost disorder free,  they are exposed to many different noises and disorders. In Ref.~\cite{Das-Sarma-Hubbard} it has been shown that the Hubbard Hamiltonian can be realized in an array of quantum dots, each hosting a single electron. Tuning the tunneling will then realize an effective spin chain model of such electron arrays.

Theoretically, preparing a many-body system to its ground state can always be achieved by cooling the system to very low temperatures, namely, below their energy gap. However, for gapless systems the energy separation between the ground state and excited states becomes vanishingly small when the size of the system increases, hence demanding for very low temperatures which are not accessible in practice. To overcome this obstacle one can use the adiabatic theorem \cite{born} according to which a many-body system always remains in the ground state of its time varying Hamiltonian  if the time variation is slow enough.  Hence, one can prepare the system in the ground state of a gapped Hamiltonian, which is practically achievable, and then changes the parameters of the system very slowly to reach the desired Hamiltonian. If this time variation is faster than the thermalization time one can guarantee then the system goes to the ground state of the desired many-body Hamiltonian. Since the energy gap of a uniform Heisenberg Hamiltonian of length $N$ goes down like $\sim 1/N$ the anti-ferromagnetic  ground state of such system is very hard to achieve.

In this paper we use the same scenario of Ref.~\cite{cirac}, proposed for optical lattices, for preparing the many-body system into its ground state. According to that proposal a series of singlet pairs, initially prepared in an optical super-lattice with alternating zero coupling, are adiabatically driven to the ground state of the uniform Heisenberg chain by switching on the couplings very slowly. In optical lattices, the main imperfection issue is the particle loss which has been studied in \cite{cirac}. Other imperfections have been analyzed in spin chain quantum communication. In \cite{petrosyan} the effect of static disorder has been investigated in an engineered XX model for perfect state transfer. The on-site energy fluctuations in spin chains have been considered in \cite{Eisert-disorder} and it was shown that these fluctuations affect the transmission in a different way compared to the static disorders. The localization problem and how to overcome it under the presence of disordered couplings and local fields have been studied in \cite{Linden-localization}

In the setup, considered here, i.e. solid state quantum dot arrays (or equivalently dopant arrays), there are very different sources of imperfection. The first one is the exchange coupling disorder resulting from imperfect fabrication and voltage gate fluctuations. Such disorders can originate from the initial fabrication process creating a random static profile for the exchange couplings or resulting from the time varying white noise  in gate voltages. The second important disorder effect is the hyperfine interaction in which the electron spin interacts with the nearby nuclei spins of the bulk. We then study the effect of such disorders in the adiabatic ground state preparation of the Heisenberg spin chain. In addition, we also introduce a mechanism for adiabatically transferring singlet-triplet states, as classical information, across such quantum chain in the same spirit of \cite{nicho} for spin qubits and of  \cite{Petrosyan-charge} for charge qubits. The difference with the ground state preparation lies in the fact that the quantum state is no longer the ground state of the Hamiltonian.  In fact, we show that while the ground state preparation is highly robust against disorder,  the singlet-triplet transfer is highly affected by such imperfections making the adiabatic strategy very inefficient for communication tasks. This justifies quantum  quenches as the more efficient way for communication across spin chains.

\begin{figure} \centering
    \includegraphics[width=8cm,height=6cm,angle=0]{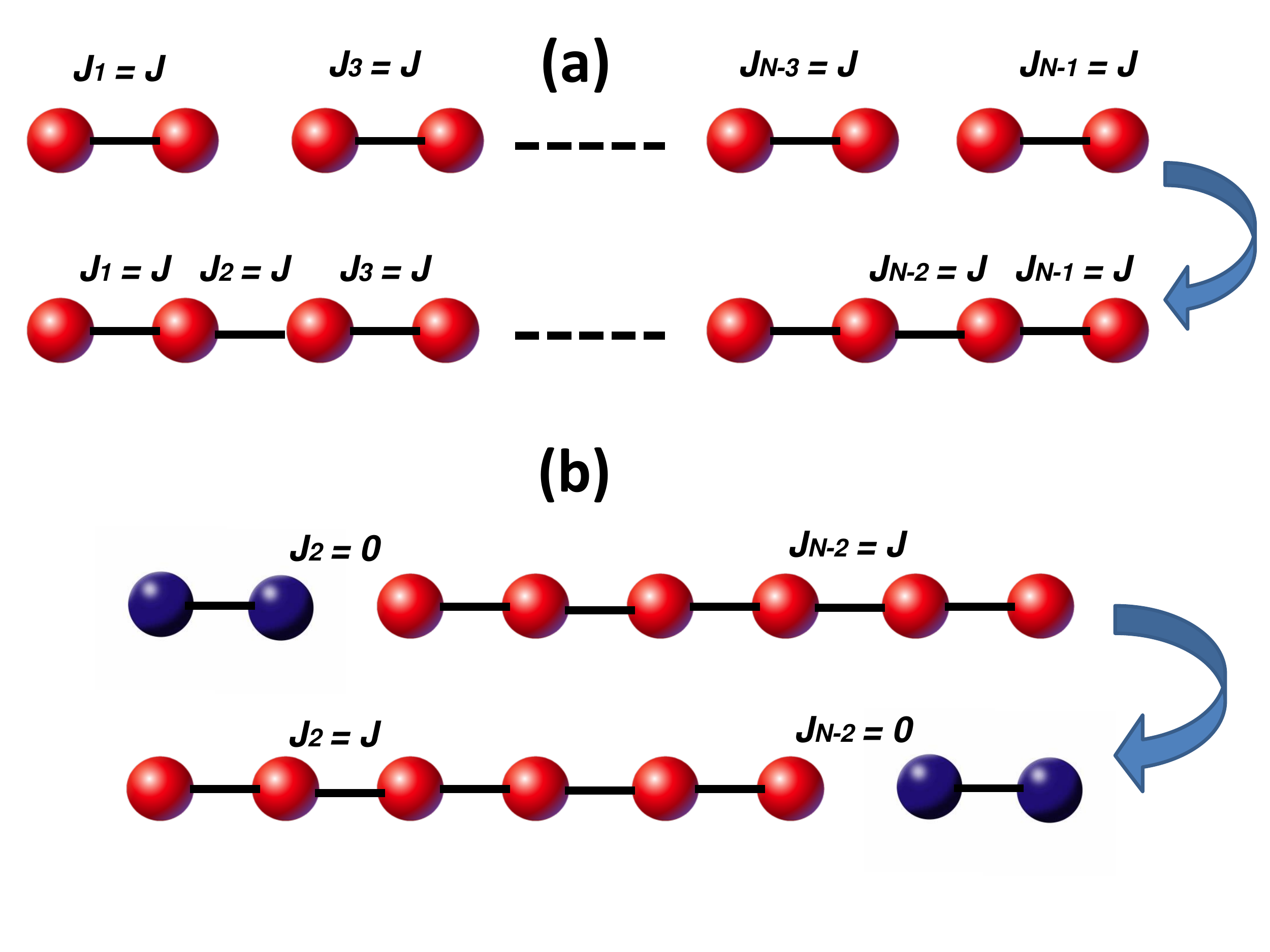}
    \caption{ (color online) (a) A fully dimerised chain, with $J_k=0$ for all even $k$, initialized in a series of singlets evolves adiabatically to the anti-ferromagnetic ground state of a uniform chain in which all the couplings are identical, i.e.  $J_k=J$. (b) A triplet (or singlet) state initially decoupled from the rest of the system is adiabatically transferred to the other side by switching on the coupling $J_2$ and switching off the coupling $J_{N-2}$ simultaneously.}
     \label{fig1}
\end{figure}

\section{Noiseless Adiabatic ground state preparation}

We consider an even chain of $N$ spin-1/2 particles (namely qubits) interacting via the Heisenberg Hamiltonian
\begin{equation} \label{Ht}
H(t)=\sum_{k=1}^{N-1} \hbar J_{k}(t) \boldsymbol{\sigma}_{k}\cdot\boldsymbol{\sigma}_{k+1},
\end{equation}
where $\boldsymbol{\sigma}_{k}=(\sigma^x_k,\sigma^y_k,\sigma^z_k)$ is the vector of Pauli operators, $J_k$ are the exchange couplings (given in Hz) and $\hbar$ is the Planck constant. The goal of our procedure is to prepare the chain in the ground state of the uniform Heisenberg model with $J_k=J$ for all $k$. Theoretically, this can be achieved by cooling the uniform Heisenberg chain via interaction with a cold reservoir whose temperature is sufficiently smaller than the energy gap of the system.  However, in practice that is notoriously difficult as the energy gap of the uniform chain goes down by increasing the length as $\sim 1/N$ and thus the needed temperatures for realizing the cooling task become unrealistic. So, to achieve that we exploit the adiabatic theorem and initialize the system into the ground state of another Hamiltonian which is easier to reach and then slowly evolve the Hamiltonian into the desired one (here the uniform Heisenberg chain). According to the adiabatic theorem if the evolution is slow enough the quantum state of the system follows the ground state manifold throughout the evolution and eventually the ground state of the desired Hamiltonian is reached. Of course, the protocol is successful only when the time needed for adiabatic evolution is less than the thermalization time.

To fulfill the above task, we assume that initially, at $t=0$, we have $J_k=J$ for all odd $k$ and $J_k=0$  otherwise. This forms a fully dimerized chain, schematically shown in Fig.~\ref{fig1}(a),  with alternating zero couplings. The ground state of such a chain is a tensor product of $N/2$ singlets as
\begin{equation} \label{singlets}
|GS(0)\ra =\bigotimes_{k=1}^{N/2} |\psi^-\ra.
\end{equation}
The first excited state of this fully dimerized Hamiltonian is highly degenerate and can be obtained by converting one of the singlets into a triplet. The energy gap for this chain is $4J$ and is independent of $N$ \cite{bathe}. Thanks to this large energy gap the initialization of the system in its ground state (\ref{singlets}) is easy and can be achieved by cooling the system into fairly low temperatures, as has been done for both cold atoms in optical lattices \cite{bloch-singlet} and electrons in double quantum dot systems \cite{Petta2005,Marcus-Cz,Yacobi-CZ}.
As time elapses, while $J_k=J$ is constant for odd $k$s, for even $k$s the exchange couplings are turned on adiabatically until they reach the value $J$ in a time $T>0$ ($T$ is the so called `ramping time'), i.e.
\begin{equation}\label{JK_t}
J_k(t)=\left\{
\begin{array}{ccc}
J & & {\rm for}\;  k \; {\rm odd} \\
\left[ t- \left(t-T\right)\theta(t-T)\right]\frac{J}{T} & & {\rm for} \; k\; {\rm even}
\end{array}\right. ,
\end{equation}
with $\theta(x)$ the step function.

We denote by $|GS(t)\rangle$ the eigenvector corresponding to the smallest eigenvalue of the operator
$H(t)$ for fixed $t$, i.e. the ground state of \eqref{Ht} at time $t$.
In order to describe the evolution of the system we employ the adiabatic theorem in that for a small, compared to
$T$, time interval $\Delta t$ (i.e. $\Delta t \ll T$) during which we can consider the Hamiltonian \eqref{Ht} as constant.\footnote{To ensure the validity of the adiabatic theorem one has to use $JT \geq (J\hbar/\Delta E)^2$, where $\Delta E $ is the energy gap of the Heisenberg Hamiltonian $H(T)$.} Then the time evolution operator in such time interval reads
\begin{equation}\label{rpu}
U(t+\Delta t, t)=\exp\left[-i H(t) \Delta t/\hbar \right].
\end{equation}
As a consequence, we determine the state of the system at any time step $j\Delta t$ ($j\in\mathbb{N}$)
by using the following relation
\begin{eqnarray}\label{rpf}
|\psi(t=j \Delta t)\rangle &=& U(j \Delta t, (j-1)\Delta t)\times  \ldots \times U(2\Delta t, \Delta t)\nonumber\\
&\times&U(\Delta t, 0) |GS(0)\rangle.
\end{eqnarray}

To see the quality of our adiabatic evolution we compute the fidelity between the quantum state of the system at time $t$  and the target state as
\begin{equation}\label{ideal}
F_g(t)=| \la  GS(T)| \psi(t)\ra| ^{2},
\end{equation}
where the subscript $g$ refers to the ground state preparation task.

\begin{figure} \centering
    \includegraphics[width=8.6cm,height=6.5cm,angle=0]{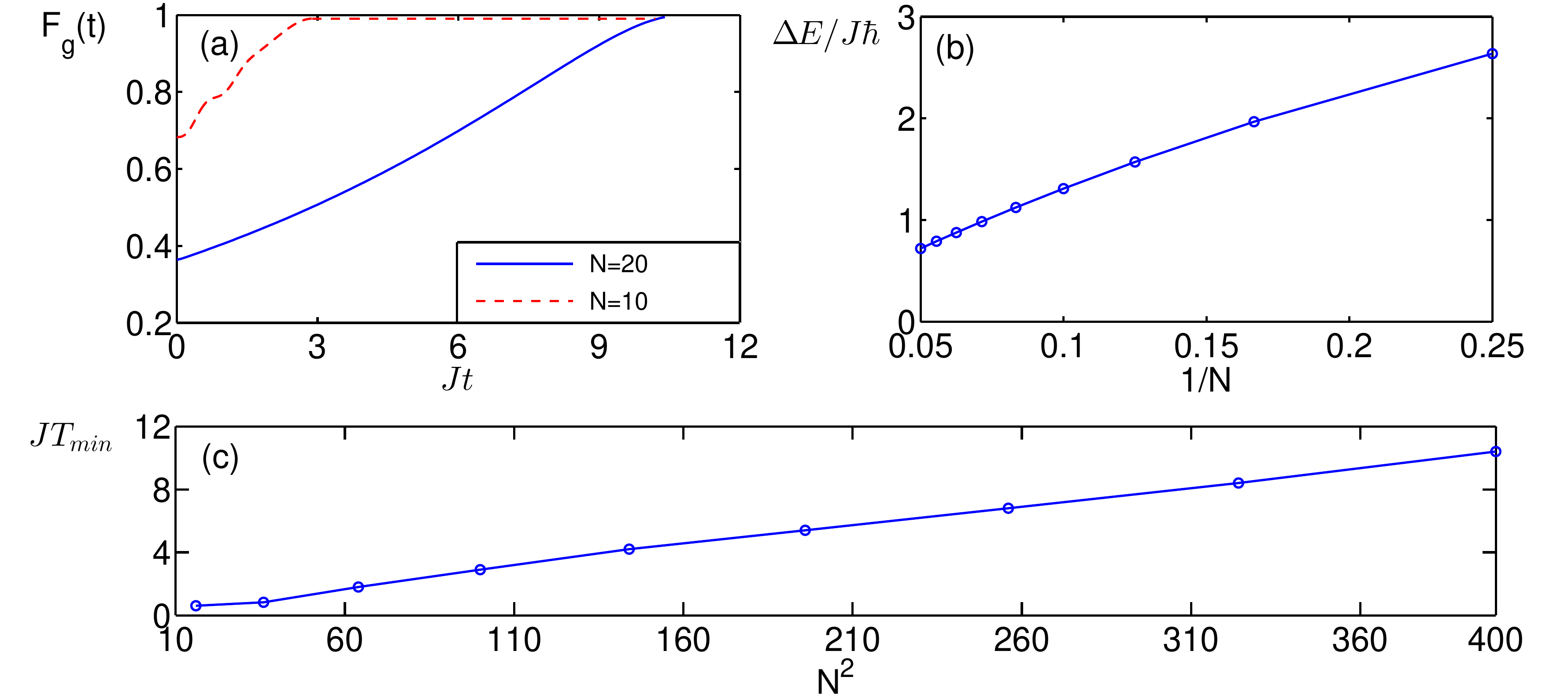}
    \caption{ (color online) (a) The ideal case of fidelity $F_g(t)$ versus  time $Jt$  for $N=10$ (using $T=2.9/J$) and $N=20$ (using $T=10.4/J$).
    (b) The Energy gap $\Delta E$ (in the units of $J\hbar$) as a function of $1/N$. (c) The ramping time $JT_{min}$ versus $N^2$. }
\label{fig2}
\end{figure}
%%%%%%%%%%%%%
In Fig.~\ref{fig2}(a) we plot $F_g(t)$ for two different lengths $N=10$ (using $T=2.9/J$) and $N=20$ (using $T=10.4/J$). As it is evident from the figure, at the end of the evolution the fidelity $F_g(T)$ is almost 1. In fact, by increasing $T$ we can always improve the final fidelity, however, it is wise to choose a high threshold such as $F_g(T)\geq 0.99$ and find the minimal ramping time $T_{min}$ which is enough to achieve such that fidelity. As mentioned above, according to the adiabatic theorem, the ramping time  $T_{min}$ is determined by the energy gap $\Delta E$. In Fig.~\ref{fig2}(b) we plot $\Delta E$, in the unites of $J\hbar$, as a function of $1/N$ which shows an almost linear dependence. This in fact suggests that
\begin{equation}\label{Tmin_N}
    JT_{min} \propto (\frac{J\hbar}{\Delta E})^2 \propto N^2.
\end{equation}
In Fig.~\ref{fig2}(c) we plot $JT_{min}$ as a function of $N^2$ which indeed confirms such dependence for large $N$. A more careful look to the numbers for $T_{min}$ shows that the adiabatic evolution for the ground state preparation is indeed very efficient and quickly drives the system into its ground state.

\section{Noiseless Adiabatic State Transfer}

We also propose to use the adiabatic switching for sending quantum information across a quantum chain. For such a scenario we assume that the information is encoded in the subspace of singlet and triplet states $|\psi^{\pm}\ra$, prepared at the beginning of the chain. The goal is to see the performance of adiabatic evolution for transferring such information. The single qubit quantum states, have already been transferred across a spin chain adiabatically \cite{nicho} and now we try to do that for a triplet (or singlet) state as well. To do so, we assume an even chain of $N$ spins with all couplings $J_k=J$ except for $J_2$ which is initially tuned to zero, decoupling the first pair of qubits from the rest of the system. A schematic picture of this configuration is shown in Fig.~\ref{fig1}(b). At $t=0$ the decoupled pair is adiabatically coupled to the system by switching on the coupling $J_2$ and simultaneously switching off the coupling $J_{N-2}$ over the time scale of $T$ as
\begin{eqnarray}\label{couplings_triplet}
    J_2&=&\left[ t- \left(t-T\right)\theta(t-T)\right]\frac{J}{T}, \cr
    J_{N-2}&=& J- \left[ t- \left(t-T\right)\theta(t-T)\right]\frac{J}{T}.
\end{eqnarray}
At the end of the process, one expects to transfer the first triplet (or singlet) pair to the last one.

To see the performance of this procedure we consider the following initial state
\begin{equation}\label{psi_0_st}
    |\Psi^\pm (0)\ra= |\psi^\pm\ra \otimes |\psi_{ch}\ra
\end{equation}
where $|\psi_{ch}\ra$ represents the ground state of a uniform Heisenberg chain with $N-2$ spins.  Using the couplings of Eq.~(\ref{couplings_triplet}) one can compute the evolution operator similarly to
Eqs.(\ref{rpu}) and (\ref{rpf}) and get the quantum state $|\Psi^\pm (t)\ra$.

The target state for transferring information is
\begin{equation}\label{psi_0_tar}
    |\Psi_{tar}^\pm\ra=  |\psi_{ch}\ra \otimes |\psi^\pm\ra
\end{equation}
in which the quantum state $|\psi^\pm\ra$ is assigned to the last pair of qubits. To quantify the quality of communication one can compute the fidelity
\begin{equation}\label{F_}
    F_c^\pm(t)=  | \la \Psi_{tar}^\pm|\Psi^\pm(t)\ra|^2,
\end{equation}
where the subscript $c$ refers to the communication task.

%%%%%%%%%%%%%%%
\begin{figure} \centering
\includegraphics[width=8.6cm,height=6.5cm,angle=0]{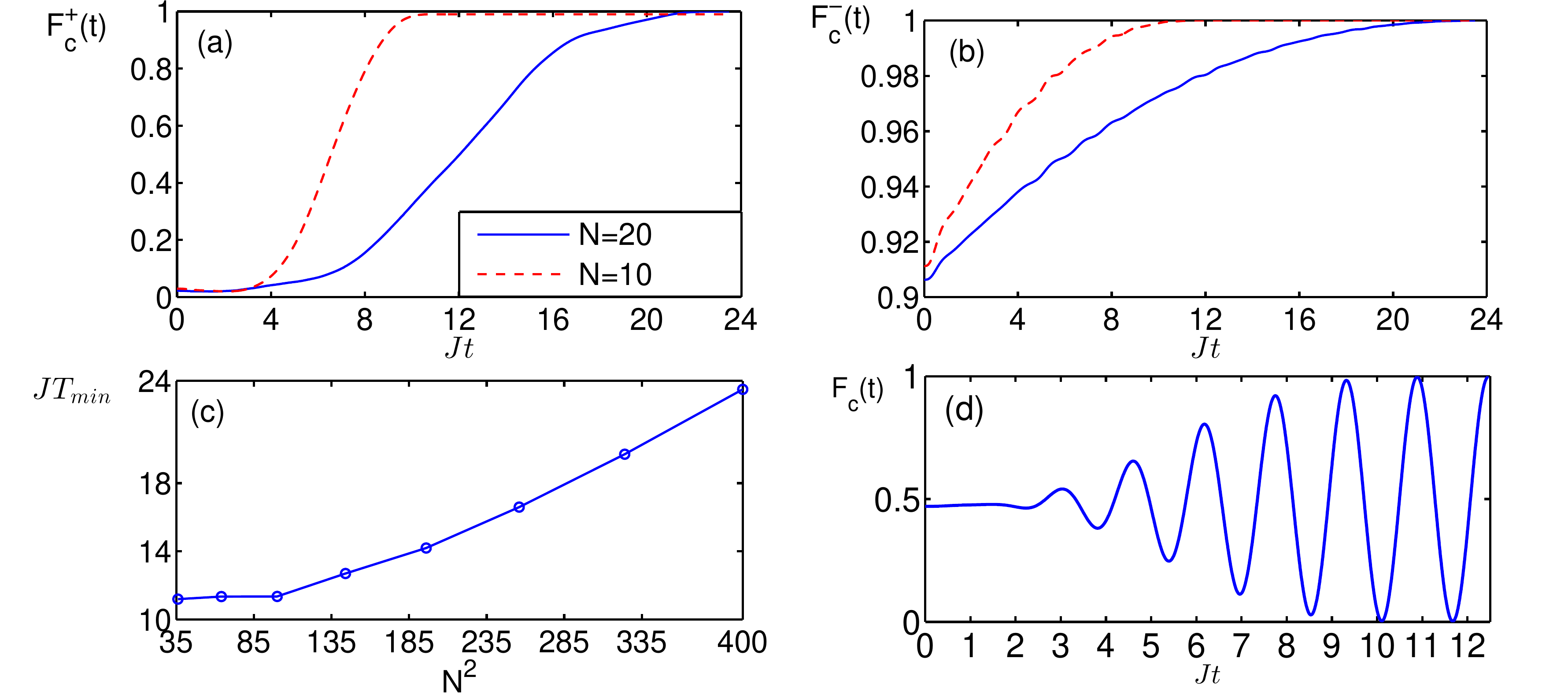}
	  \caption{ (color online)
	  (a) The ideal case of fidelity $F_{c}^{+}(t)$ versus time $Jt$  for $N=10$ (using $T=11.36/J$) and $N=20$ (using $T=23.5/J$).
    (b) The ideal case of fidelity $F_{c}^{-}(t)$ versus time $Jt$  for $N=10$ (using $T=11.36/J$) and $N=20$ (using $T=23.5/J$). (c) The minimum ramping time for singlet-triplet state transfer $JT_{min}$ versus $N^2$. (d) The ideal case of fidelity $F_c(t)$ for an equally weighted superposition of singlet and triplet versus  $Jt$ for $N=10$ (using $T=11.36/J$). }
\label{fig3}
\end{figure}
%%%%%%%%%%%%%%%

In Figs.~\ref{fig3}(a) and (b) we plot the $F_c^+(t)$ and $F_c^-(t)$ as a function of time respectively for two different chains of length $N=10$ (choosing $T=11.36/J$) and $N=20$ (choosing $T=23.5/J$). Just as before, choosing the threshold $F_c^\pm(T)>0.99$ determines the minimal ramping time $JT_{min}$ for any $N$ to achieve such a fidelity. Notice that  such a time $JT_{min}$ is the same for singlet and triplet state as shown in Figs.~\ref{fig3}(a and b).
In Fig.~\ref{fig3}(c) the ramping time $JT_{min}$ needed for singlet-triplet state transfer is shown versus $N^2$ which clearly shows a linear dependence for larger chains. A simple comparison between the minimum ramping time $JT_{min}$ for state transfer (see Fig.~\ref{fig3}(c)) and the ground state preparation (see Fig.~\ref{fig2}(c)) shows that the the adiabatic communication scheme is at least 3 times slower than the ground state preparation. This is due to the fact that these two evolutions happen in two different subspaces, namely the ground state preparation takes place in the ground state manifold while the communication mechanism takes place in the excited state subspace. The energy separation between the quantum state with one triplet pair and the relevant higher energy states is lower than the energy separation in the global singlet subspace, needed for ground state preparation. This indeed shows itself in the larger time scales needed for accomplishing these two different tasks.

Finally, it is worth saying that any superposition of singlet and triplet can be perfectly transmitted. However, due to the fact that triplet and singlet have different energies they get a relative phase in time, even when the dynamics has not yet started, which has to be taken to account. This is equivalent to a deterministic rotation around the $z$-axis in the Bloch sphere of the qubit which has to be considered before any further computational operations.
As result the fidelity, achieved by the adiabatic evolution, becomes oscillatory and reaches one on its maxima. As an example we consider an equally weighted superposition $(|\psi^-\rangle+|\psi^+\rangle)/\sqrt{2}$ and show the corresponding fidelity versus time in Fig. \ref{fig3}(d) in which the oscillations in fidelity continues even when the adiabatic ramping is finished.

\section{Imperfections}

The above procedures are of course very ideal and in realistic scenarios one may expect to have disorder in the Hamiltonian which deteriorates the quality of the protocol. In this section we consider the most common disorder effects in solid state realization of spin chains, namely, static and time varying disorder in exchange couplings and hyperfine interaction with surrounding nuclei spins on the performance of both protocols.

\subsection{Disordered exchange couplings}

In a typical array of quantum dots \cite{Smith-QDarray} each loaded with single electrons one can control the exchange interaction using electric gate voltages \cite{Petta2005,Marcus-Cz,Yacobi-CZ}. However, the imperfect fabrications and the inevitable gate voltage fluctuations will introduce disorder in the exchange couplings. The induced disorder can be classified in to two different categories: (i) static disorder mainly because of imperfect fabrications and possible impurities in the system and; (ii) time varying fluctuations resulting from voltage fluctuations of the gates.

We assume that the disorders emerging on each coupling is independent of the others and can be simulated as \cite{doherty}
\begin{equation} \label{J_random}
J_k \rightarrow J_k e^{-\epsilon^{(k)}(t)}
\end{equation}
with
\begin{equation}\label{exch}
\epsilon^{(k)}(t)=\epsilon^{(k)}_{static}+\epsilon^{(k)}_{white}(t),
\end{equation}
where $\epsilon^{(k)}_{static}$ is a random number with uniform distribution in the interval $[-\delta, \delta]$, for some constant number $\delta$ and $\epsilon^{(k)}_{white}(t)$	is the white noise with the frequency spectrum $\eta$ ($\delta$ and $\eta$ are dimensionless quantities).

The phenomenological fit in Eq.~(\ref{J_random}) can faithfully explain the coupling fluctuations in GaAs singlet-triplet qubit experiments \cite{Yacoby-PRL-2013,Yacoby_NatPhys-2009}. In practice, due to the strong cross-capacitive couplings between electrostatic gates in arrays of dots the coupling fluctuations  between different pairs of electrons might be correlated. However, the most destructive scenario is the independent fluctuations of the form of Eq.~(\ref{J_random}) as, for instance, the perfect correlation between all exchange coupling fluctuations implies that $\epsilon^{(k)}(t)$ becomes independent of site $k$ and all couplings fluctuate in the same way. This simply means that all the energy levels fluctuate together and its impact will be an irrelevant global phase with no destructive effect at all. On the other hand, independent fluctuations, considered in Eq.~(\ref{J_random}) maximizes the relative energy fluctuations between any pair of energy levels and thus makes the worst scenario which we consider in this paper.

\subsection{Hyperfine interaction}

For electron spins in quantum dots, the most destructive phenomenon is interaction with the spin of nuclei in the bulk, i.e., hyperfine
interaction. In a solid state hetero-structure, such as GaAs gated quantum dots, the electron spin interacts with many nuclear
spins of its host material, and it can be described as  $H_{HF}=\sum_{j=1}^{M}a_{j} {\textbf{I}}_{j}\cdot\boldsymbol{\sigma},$ in which ${\textbf{I}}_{j}$ denotes the spin of the $j_{th}$ nucleus, $\boldsymbol{\sigma}$ is the Pauli vector operator representing the electron spin, and $a_{j}$ represents the coupling strength between the $j$'th nucleus and the electron spin. Due to the very slow dynamics of nuclei spins in comparison to the time scales of our protocol one can describe the average effect of nuclear spins as effective magnetic field ${\textbf{B}}$,	such that: $(\sum_{j=1}^{M}a_{j} {\textbf{I}}_{j})\cdot\boldsymbol{\sigma}={\textbf{B}}\cdot\boldsymbol{\sigma}$. Incorporating the hyperfine interaction modifies the Hamiltonian $H(t)$, given in Eq.~(\ref{Ht}), as
\begin{equation} \label{H_hyperfine}
H(t) \rightarrow H(t)=\sum_{k=1}^{N-1} \hbar J_{k}(t) \boldsymbol{\sigma}_{k}\cdot\boldsymbol{\sigma}_{k+1} + \sum_{k=1}^{N}{\textbf{B}}_{k}\cdot\boldsymbol{\sigma}_{k},
\end{equation}
where the nuclear field ${\textbf{B}}_{k}$ is a three-dimensional random vector. Under the quasi-static approximation the spin of nuclei do not change
in the state transferring time scale and ${\textbf{B}}_{k}$ is supposed
to be time independent. In the large $M$ limit, the amplitude of the vectors ${\textbf{B}}_{k}$ have a Gaussian distribution
\begin{equation}\label{bnuc}
P({\textbf{B}})=\frac{1}{(2\pi B_{nuc}^{2})^{3/2}}\exp\left(-\frac{ {\textbf{B}}\cdot {\textbf{B}}} { 2B_{nuc}^{2}}\right),
\end{equation}
in which the $B_{nuc}$ is the variance of the distribution.
	
%%%%%%%%%%%%%
\begin{figure} \centering
    \includegraphics[width=8.6cm,height=6.5cm,angle=0]{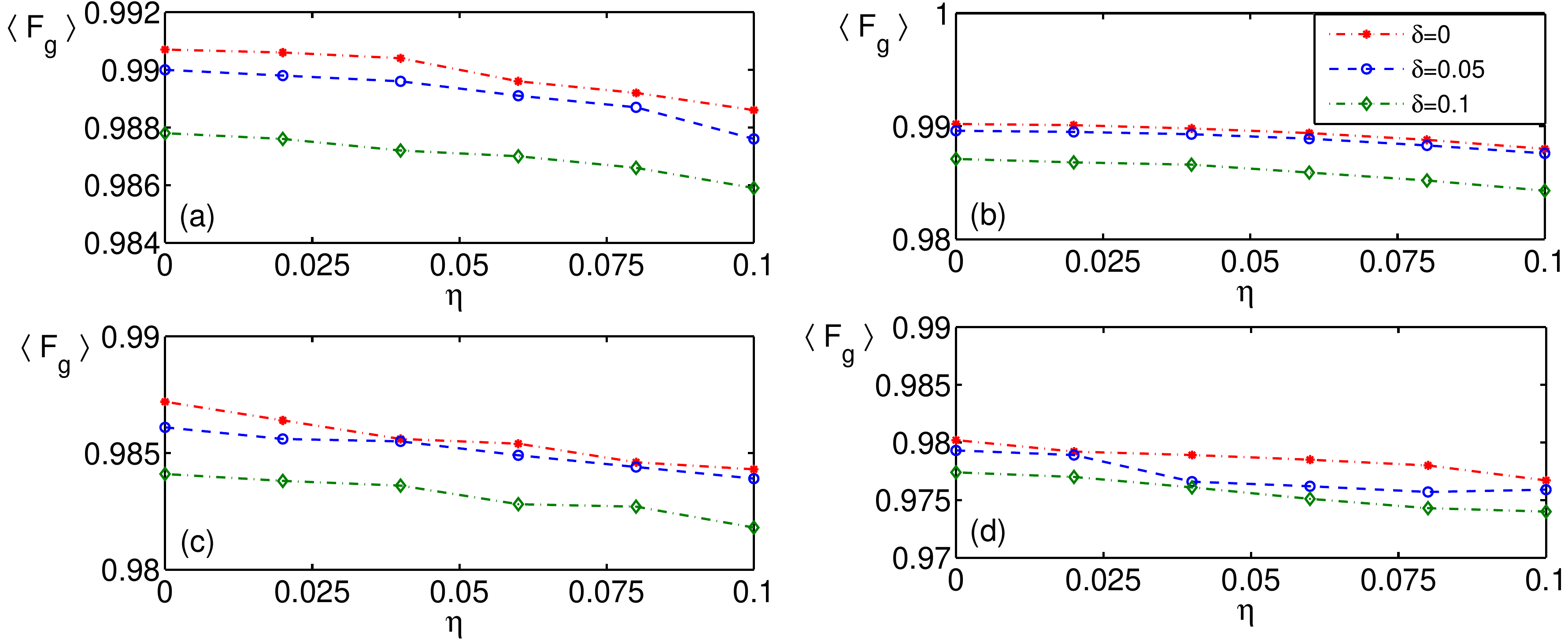}
    \caption{ (color online) The average ground state fidelity $\la F_g \ra$ at $JT=2.9$ versus  $\eta$
	  (varience of white noise) for different static noise $\delta$  and with different hyperfine interaction: (a) $B_{nuc}=0$; (b) $B_{nuc}=0.02$; (c) $B_{nuc}=0.06$; (d) $B_{nuc}=0.1$.}
     \label{fig4}
\end{figure}
%%%%%%%%%%%%%

%%%%%%%%%%%%%%%%%%%%%%%%%%%%%%%%%%%%%%%%%%%%%%%%%%%%%%%%

\section{Results}

In this section we consider the effect of disorder in exchange couplings together with hyperfine interaction on the performance of both adiabatic ground state preparation and adiabatic singlet-triplet communication. In the presence of disorder in order to have meaningful quantities we repeat our protocol (section IV(A,B)) for 100 different realization and make an average over all realizations for the ground state fidelity $F_g(T_{min})$
(represented by $\langle F_g \rangle$ and the
singlet-triplet communication fidelity $F_c^\pm(T_{min})$ (denoted by $\langle F_c^\pm \rangle$).
For each realization we choose a random set of numbers $\epsilon^{(k)}_{static}$ (for $k=1,2,...,N-1$) uniformly distributed in $[-\delta,+\delta]$ for any fixed parameter $\delta$ and similarly produce a set of random magnetic fields ${\textbf{B}}_{k}$ according to the normal distribution (\ref{bnuc}). The white noise term $\epsilon^{(k)}_{white}(t)$ is generated using the method of Ref.~\cite{Lennon-WhiteNoise} (see also \cite{bayat-song-QD}) and varies at each time step during the time integration.

In Figs.~\ref{fig4}(a)-(d) we plot the ground state fidelity $\la F_g \ra$ as a function of white noise strength $\eta$ for different static noise power $\delta$ and hyperfine interaction $B_{nuc}$ in a chain of length $N=10$. Indeed, the results show that the ground state preparation is very robust against all kind of disorders as the fidelity $\la F_g \ra$  does not go below $0.97$ even for strong disorders.

%%%%%%%%%%%%%
\begin{figure} \centering
\includegraphics[width=8.6cm,height=6.5cm,angle=0]{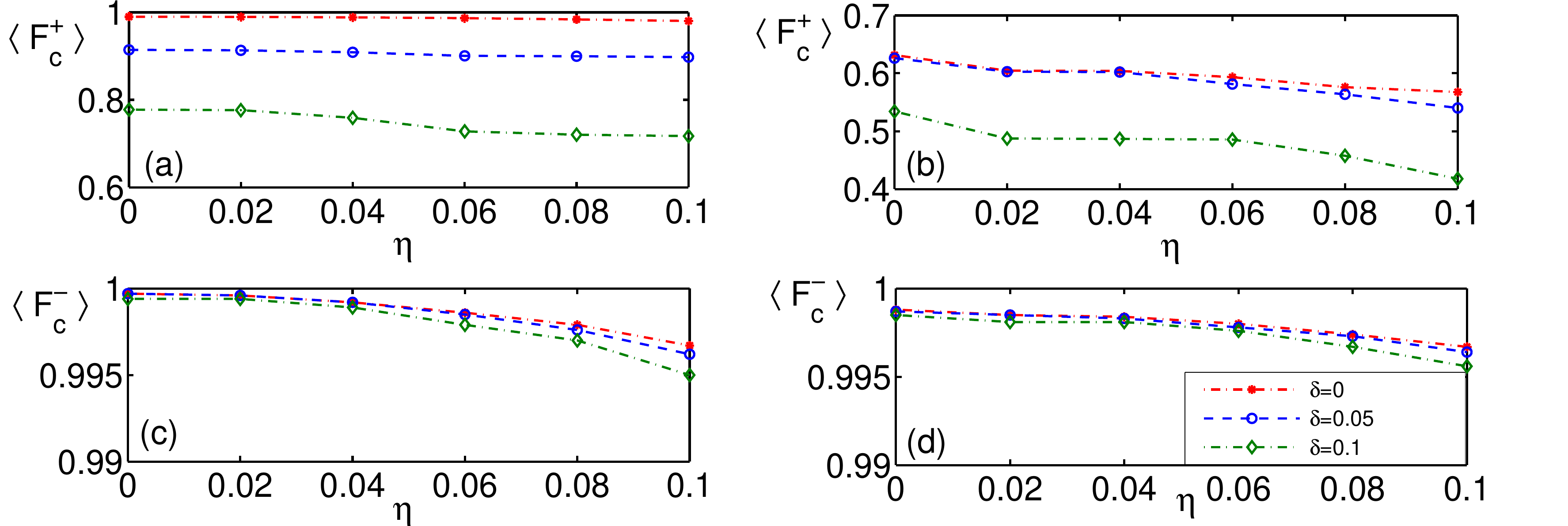}
	  \caption{  (color online) The average  fidelity for triplet transfer $\la F_c^+ \ra$ and singlet transfer $\la F_c^- \ra$ at $JT=11.36$ versus $\eta$ (variance of white noise) for different static noise $\delta$ and  with different hyperfine interaction:  (a) and (c\,) $B_{nuc}=0$; (b) and (d) $B_{nuc}=0.1$. }
\label{fig5}
\end{figure}
%%%%%%%%%%%%%

In Fig.~\ref{fig5} we show the results for singlet-triplet quantum communication. Actually we plot
$\la F_c^{\pm} \ra$  versus white noise strength $\eta$ for different values of static noise power $\delta$ and hyperfine interaction $B_{nuc}$. In comparison to the ground state preparation the communication fidelity is very susceptible to disorder for triplet state transfer as for instance for $B_{nuc}=0.1$  and $\eta=\delta=0.1$ the fidelity goes down to $\la F_c^+ \ra=0.42$. In contrast, the communication fidelity for singlet state transfer is quite robust even in the presence of strong disorder the fidelity does not go below $0.995$. This is due to the different dimension of the triplet and singlet subspaces where the evolution is taking place. Furthermore, one can see that the white noise fluctuation has less effect on $\la F_c^{\pm} \ra$ in comparison to the static noise and hyperfine interaction. This is due to the fact that the time varying fluctuations cancel each other over time and thus create less effect on the performance of the system.
The most destructive effect of all disorders can be seen for hyperfine interaction since even in the absence of all other noises (i.e. $\eta=\delta=0$) the hyperfine noise of $B_{nuc}=0.1$ gives a very low fidelity of $\la F_c^+ \ra=0.54$. This is because the random magnetic field generated by the nuclei spins have random direction and thus change the total magnetization of the system during the evolution while the exchange coupling disorder preserves the total magnetization of the system.

\begin{figure} \centering
\includegraphics[width=9cm,height=4.5cm,angle=0]{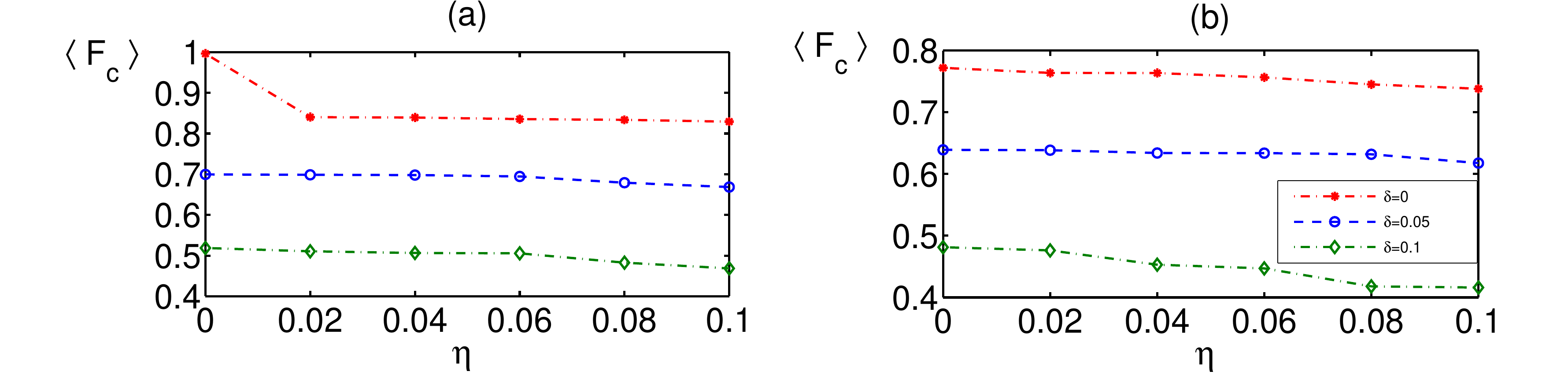}
	  \caption{ (color online) The average fidelity for transferring an equally weighted superposition of singlet and triplet at $JT=11.36$ versus $\eta$ (variance of white noise) for different static noise $\delta$ and  with different hyperfine interaction:  (a) $B_{nuc}=0$; (b) $B_{nuc}=0.1$. Since there is oscillations in the attainable fidelity the time that fidelity peaks is slightly after the ramping time $T$, namely $t=12.45/J$.}
\label{fig6}
\end{figure}

Finally, we can say that the noise has a destructive impact as soon as we go outside the singlet subspace,
 in fact already in a superposition like $(|\psi^-\rangle+|\psi^+\rangle)/\sqrt{2}$ the noise causes the fidelity falling down to quite small values (see Figs. \ref{fig6}(a)-(b)) .

%%%%%%%%%%%%%
%%%%%%%%%%%%%

%%%%%%%%%%%%%%%%%%%%%%%%%%%%%%%%%%%%%%%%%%%%%%%%%%%%%%%%

\section{Time Scales and Limitations}

In recent experiments the exchange coupling $J \simeq 0.5-1$ GHz has been experimentally realized \cite{Yacobi-CZ,Marcus-Cz}.  According to the data shown in Fig.~\ref{fig2}(c) the minimum time needed for the preparation of the ground state of a chain of $N=20$ electrons, starting from ten pairs of singlets, is $JT_{min} \simeq 10$. Using the experimental values of $J$, one can see that the time needed for such initialization varies between $10$ to $20$ ns. The same estimation for the minimum time needed for state transfer, given in Fig.~\ref{fig3}(c), across a chain of length $N=20$ will be between $25$ to $50$ ns.

As discussed before the minimum time needed for our adiabatic processes increases as $N^2$. The main limiting factors for the number of electrons are the coherence time of the system which is determined  by the hyperfine interaction and the thermalization induced by the finite temperature of the dilution fridges. The coherence times of $\sim 50$ ns \cite{Marcus-Cz} to $500$ ns \cite{Yacobi-CZ} have been observed for two pairs of singlet-triplet electronic qubits in coupled quantum dots. This can be significantly improved to $\sim 200$ $\mu$s using spin echo pulses \cite{Yacoby-NatPhys-2011}. Taking a pessimistic coherence time of $50$ ns and an exchange coupling of $J=0.5$ GHz shows that our protocol for ground state preparation can be still effective for chains up to length $N\approx 40$.

Determining the thermalization time due to the finite temperature ($\sim 50$ mK) of the dilution fridges  is more tricky as unlike the coherence time there is not much experimental data. According to Fig.~\ref{fig2}(b) the energy gap of a chain of length $N=20$ is $\Delta E \simeq J\hbar$. In a quantum dot array with the typical exchange coupling of $J  \simeq 0.5$ GHz \cite{Marcus-Cz} this energy gap is equivalent to $\sim 5$ mK, one order of magnitude smaller than the typical temperatures of normal dilution refrigerators. This means that simple cooling cannot cool the system into its ground state and indeed that is why we proposed our adiabatic evolution mechanism for preparation of the ground state. To see up to what time scales the system can still faithfully stays in its ground state after a adiabatic evolution needs more experimental investigations as at this stage there are not much information about the thermalization time of the electronic spins in quantum dot arrays. We believe that, very likely, this time scale is not faster than the coherence time of the system and thus one still can operate our mechanism for chains up to $N\leq 40$. \\

%%%%%%%%%%%%%%%%%%%%%%%%%%%%%%%%%%%%%%%%%%%%%%%%%%%%%%%%

\section{Conclusion}

In summary, we have considered the effect of two inevitable types of disorder, namely hyperfine interaction and exchange coupling fluctuation,
in quantum dot arrays for adiabatically preparation of ground state and singlet-triplet state transfer. The ground state preparation, performed in the ground state manifold of the Hamiltonian during the time evolution, is accomplished in much faster time scales in comparison to the singlet-triplet communication which is operated on the excited state subspace. Moreover, our analysis shows that the ground state preparation is highly robust against disorder and the performance remains excellent even in the presence of strong disorders. On the other hand, the adiabatic communication scheme shows relatively poor performance in the presence of disorders. In particular, the hyperfine interaction deteriorates the fidelity very strongly as such interaction with nearby nuclear spins does not preserve the magnetization during the evolution.

The main consequence of this is that while in a quantum dot array the adiabatic strategy is very efficient for preparing a many-body system in its ground system, but it is not much reliable for quantum states transfer.
For such a task it may be better to use non-adiabatic evolution or to switch to quantum  quenches  \cite{bose}.

\section{Acknowledgements}
UF would like to thank the University College London for the kind hospitality.
AB and SB  acknowledge support from
EPSRC grant EP/K004077/1 (nano-electronic based quantum
technologies), SB also thanks the ERC grant PACOMANEDIA.

\end{document}